\documentclass{aa}

\usepackage{graphicx}
\begin{document}

   \title{The bends in the slopes of radial abundance gradients
   in the disks of spiral galaxies -- do they exist?}

\author{L.S.~Pilyugin \inst{1} }

  \offprints{L.S. Pilyugin }

   \institute{   Main Astronomical Observatory
                 of National Academy of Sciences of Ukraine,
                 27 Zabolotnogo str., 03680 Kiev, Ukraine, \\
                 (pilyugin@mao.kiev.ua)
                 }

\date{Received 14 March 2002 / accepted 1 October 2002}

\abstract{
Spiral galaxies with a reported bend in the slope of gradient in the oxygen 
abundances (O/H)$_{R_{23}}$, derived with traditionally used R$_{23}$ -- method, 
were examined. It is shown that the artificial origin of the reported bends can 
be naturally explained. Two situations that result in a false bend in the slope 
of (O/H)$_{R_{23}}$ gradient are indicated. It is concluded that at the present 
time there is no example of a galaxy with an undisputable established bend in 
the slope of the oxygen abundance gradient.
}

\titlerunning{bends in the slopes of abundance gradients}

\authorrunning{L.S.~Pilyugin}

\maketitle

\section{Introduction}

It has been known for a long time (Searle 1971, Smith 1975) that disks of spiral
galaxies can show radial oxygen abundance gradients, in the sense that
the oxygen abundance is higher at the central part of disk and decreases
with galactocentric distance. By now, spectra have been obtained for hundreds
of H\,{\sc ii} regions in disks of spiral galaxies.
The characteristic oxygen abundances (the oxygen abundance at a predetermined
galactocentric distance) and radial oxygen abundance gradients were obtained for
a large sample of spiral galaxies (Vila-Costas \& Edmunds 1992, Zaritsky et al.
1994, van Zee et al. 1998, among others).
It was obtained that nearly all the gradients are reasonably well fitted by a
single exponential profile, although in several cases the gradient slope may
not be constant across the disk but instead flattens (or steepens) in the
outer disk.

Zaritsky (1992) has hypothesized that the oxygen abundance gradients in the disks
of spiral galaxies flatten noticeably at the radius where the rotation curve
changes from linearly rising to flat, and he has suggested a star-forming
viscous disk model of galaxy evolution to argue in favour of his hypotheses.
Friedli et al. (1994) and Friedli \& Benz (1995) have predicted breaks in
the slope of abundance gradients in spiral galaxies with young bars
($<$ 0.5$\div$1~Gyr). The break in the slope of abundance gradient in spiral 
galaxy then was considered as an indicator of a recently formed bar in that 
galaxy (Roy \& Walsh 1997, Friedli 1999, Consid\'ere et al. 2000).

The signs of a break in the slope of abundance gradients were found on the basis  
of oxygen abundances derived with the abundance indicator R$_{23}$ or/and 
R$_{3}$ (R$_{23}$ -- method) (Vilchez et al. 1988, Zaritsky 1992, Vila-Costas \& 
Edmunds 1992, Zaritsky et al. 1994, Roy \& Walsh 1997). Zaritsky (1992) and 
Zaritsky et al. (1994) noted that the bend in the abundance indicator 
R$_{23}$ (or R$_{3}$) gradient may not reflect a corresponding bend in the 
abundance gradient. Recently it has been shown (Pilyugin 2000, 2001a,b) that the 
oxygen abundance derived with the R$_{23}$ -- method involves a systematic error
depending on the excitation parameter P: the R$_{23}$ -- method provides more
or less realistic oxygen abundances in high-excitation H\,{\sc ii} regions and
yields an overestimated oxygen abundances in low-excitation H\,{\sc ii} regions.
This is in agreement with the result of Kinkel \& Rosa
(1994), who showed the need to lower all H\,{\sc ii} region abundances
obtained on the basis of the R$_{23}$ calibration after Edmunds and Pagel (1984)
at intrinsic solar-like O/H values and above. Castellanos et al. (2002) also
found that the R$_{23}$ -- method yields an overestimated oxygen abundance 
in low-excitation H\,{\sc ii} regions.  A new way of oxygen abundance
determination in H\,{\sc ii} regions (P -- method) was suggested
(Pilyugin 2001a). It was demonstrated that the oxygen abundances derived with
the P -- method are as credible as ones derived with the classic T$_{e}$ --
method (Pilyugin 2001a,b). It should be noted, however, that the P -- method 
is established on the basis of H\,{\sc ii} regions with R$_{23}$ larger than 2 
and the validity of this method in the case of H\,{\sc ii} regions with R$_{23}$ 
less than 2 (most metal-rich H\,{\sc ii}  regions) may be disputed. 

The radial distribution of oxygen abundances derived through the P -- method
in some galaxies with a reported bend in the slope of abundance will be considered 
here to examine the reality of the reported bends.

\section{Galaxies with a reported bend in the slope of the abundance gradient}

\subsection{Preliminary remarks}

The following relationship between oxygen abundance and strong line intensities
was suggested in Pilyugin (2001a)
\begin{equation}
12+log(O/H)_{P} = \frac{R_{23} + 54.2  + 59.45 P + 7.31 P^{2}}
                       {6.07  + 6.71 P + 0.37 P^{2} + 0.243 R_{23}}  ,
\label{equation:ohp}
\end{equation}
where $R_{23}$ =$R_{2}$ + $R_{3}$,
$R_{2}$ = $I_{[OII] \lambda 3727+ \lambda 3729} /I_{H\beta }$,
$R_{3}$ = $I_{[OIII] \lambda 4959+ \lambda 5007} /I_{H\beta }$,
P = $R_{3}$/$R_{23}$.
The oxygen abundance in the H\,{\sc ii} region derived with Eq.(\ref{equation:ohp}) will be
referred to as (O/H)$_{P}$.
The relationship between oxygen abundance and strong line intensities has two
values with two distinctive parts named usually the lower and upper branches
of the R$_{23}$ -- O/H diagram, and so one has to know in advance on which
branch of the R$_{23}$ -- O/H diagram the H\,{\sc ii} region lies. The above
expression for the oxygen abundance determination in H\,{\sc ii} regions,
Eq.(\ref{equation:ohp}), is valid for H\,{\sc ii} regions with 12+log(O/H) higher
than around 8.2.

The O/H -- R$_{23}$ calibration after Edmunds \& Pagel (1984) will be used
here for the oxygen abundance determination with the R$_{23}$ -- method.
For the sake of convenience, their calibration for the upper branch
of the R$_{23}$ -- O/H diagram was approximated by the polynomial
\begin{eqnarray}
12+log(O/H)_{R_{23}} = 9.302 - 0.403x - 0.675x^2
\nonumber  \\
- 0.701x^3 + 0.666x^4
\label{equation:ohep}
\end{eqnarray}
where $x = \log R_{23}$.
The oxygen abundance in H\,{\sc ii} region derived with Eq.(\ref{equation:ohep}) will be
referred to as (O/H)$_{R_{23}}$.

\subsection{False bends of the "first type"}

We will start from a consideretion of the well-observed galaxy M101. 
The break in the slope of the abundance gradient in M101
was reported by Zaritsky (1992) and by Vila-Costas \& Edmunds (1992).
It should be noted that this conclusion was not confirmed by 
Henry \& Howard (1995), who determined oxygen abundance using a sequential
photoionization model analysis procedure and found that the observed behavior
of line ratios accross the disk of M101 is consistent with an abundance gradient
that is exponential in form and has a constant slope. About ninety measurements 
of the strong oxygen lines intensities in H\,{\sc ii} regions of the galaxy M101 
were published (Garnett \&  Kennicutt 1994, Garnett et al. 1999, Kennicutt \&  
Garnett 1996, Kinkel \&  Rosa 1994, McCall et al. 1985, Rayo et al. 1982,
Searle 1971, Shields \& Searle 1978, Smith 1975, Torres-Peimbert et al. 1989,
van Zee et al. 1998).

\begin{figure}
\resizebox{0.80\hsize}{!}{\includegraphics[angle=0]{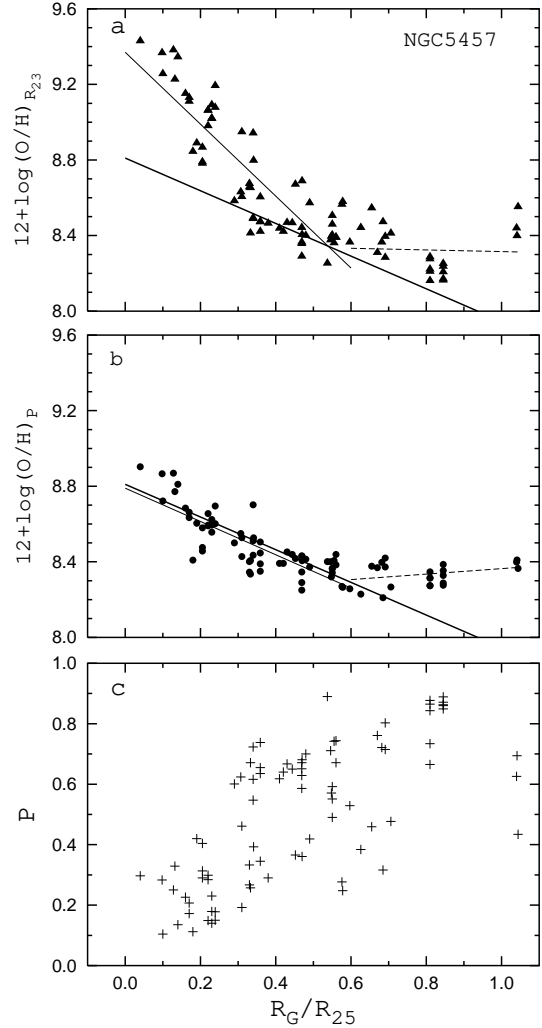}}
\caption{
Gradients in the properties of M101. The galactocentric distances are normalized
to the isophotal radius R$_{25}$.
{\bf a)} The triangles are (O/H)$_{R_{23}}$ abundances in H\,{\sc ii} regions derived
with Eq.(\ref{equation:ohep}).
The thin solid line is the best fit to the H\,{\sc ii} regions with galactocentric distances
less than 0.6R$_{25}$, the dashed line is the best fit to the H\,{\sc ii} regions with
galactocentric distances larger than 0.6R$_{25}$.
The thick solid line is the the (O/H)$_{T_{e}}$ --  R$_G$ relation obtained in
Pilyugin (2001b).
{\bf b)} The filled circles are (O/H)$_{P}$ abundances in the same H\,{\sc ii} regions
derived with Eq.(\ref{equation:ohp}).
The thin solid line is the best fit to to H\,{\sc ii} regions with galactocentric distances
less than 0.6R$_{25}$, the dashed line is the best fit to H\,{\sc ii} regions with
galactocentric distances larger than 0.6R$_{25}$.
The thick solid line is the the (O/H)$_{T_{e}}$ --  R$_G$ relation.
{\bf c)} The excitation parameter P as a function of galactocentric distance.
}
\label{figure:izl5457}
\end{figure}

\begin{figure}
\resizebox{0.80\hsize}{!}{\includegraphics[angle=0]{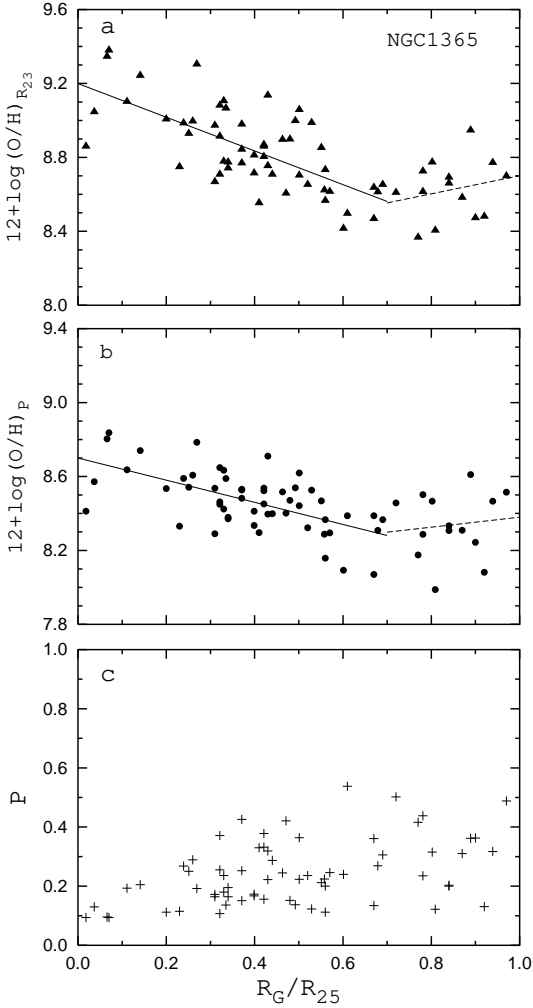}}
\caption{
Gradients in the properties of NGC1365. The galactocentric distances are
normalized to the isophotal radius R$_{25}$.
{\bf a)} The triangles are (O/H)$_{R_{23}}$ abundances in H\,{\sc ii} regions derived
with Eq.(\ref{equation:ohep}).
The thin solid line is the best fit to the H\,{\sc ii} regions with galactocentric distances
less than 0.7R$_{25}$, the dashed line is the best fit to the H\,{\sc ii} regions with
galactocentric distances larger than 0.7R$_{25}$.
{\bf b)} The filled circles are (O/H)$_{P}$ abundances in the same H\,{\sc ii} regions
derived with Eq.(\ref{equation:ohp}).
The thin solid line is the best fit to to H\,{\sc ii} regions with galactocentric distances
less than 0.7R$_{25}$, the dashed line is the best fit to H\,{\sc ii} regions with
galactocentric distances larger than 0.7R$_{25}$.
{\bf c)} The excitation parameter P as a function of galactocentric distance.
}
\label{figure:izl1365}
\end{figure}

Fig.\ref{figure:izl5457}a shows the radial (O/H)$_{R_{23}}$ abundance
distribution  (the triangles) for the H\,{\sc ii} regions in M101  derived
with Eq.(\ref{equation:ohep}). The galactocentric distances are
normalized to the isophotal radius R$_{25}$ which was taken to be equal to
14.42 arcmin (de Vaucouleurs et al. 1991). Fig.\ref{figure:izl5457}b shows the 
radial (O/H)$_{P}$ abundance distribution  (the filled circles) for H\,{\sc ii} 
regions in M101  derived with Eq.(\ref{equation:ohp}) using the same values 
of the strong oxygen lines intensities. 
As can be seen in Fig.\ref{figure:izl5457}a,b there is break in the slope
of (O/H)$_{R_{23}}$ abundance distribution as in the slope of (O/H)$_{P}$
abundance distribution, at the galactocentric distance
R$^*$ $\sim$ 0.6R$_{25}$. Close examination of Fig.\ref{figure:izl5457}a,b shows
that both the (O/H)$_{R_{23}}$ and the (O/H)$_{P}$ distributions in the disk
of M101 flatten at the radius R$^*$ where the oxygen abundance decreases 
to around 12+log(O/H)
= 8.2. The oxygen abundances in H\,{\sc ii} regions with galactocentric
distances larger than R$^*$ are expected to be less than
12+logO/H=8.2, i.e they do not belong to the upper branch of the R$_{23}$ -- O/H
diagram. But Eq.(\ref{equation:ohp}) and Eq.(\ref{equation:ohep})
can be used for oxygen abundance determination in H\,{\sc ii} regions of the
upper branch of the R$_{23}$ -- O/H diagram only. Then one can conclude that 
the use of these equations for the oxygen abundance determination in the 
H\,{\sc ii} regions
of M101 with galactocentric distances larger than R$^*$ results in the
wrong (O/H)$_{R_{23}}$ and (O/H)$_{P}$ abundances in these H\,{\sc ii} regions and,
as cosequence, results in the false breaks in the slopes
of (O/H)$_{R_{23}}$ and (O/H)$_{P}$ abundance distributions.
In the case of M101 this problem was clearly noted by Kennicutt \& Garnett (1996). 
Since they did not use the R$_{23}$ calibration for the outermost H\,{\sc ii} regions, 
their abundance gradient does not show the break.

Around twenty individual measurements of the temperature-sensitive line ratios
in H\,{\sc ii} regions of M101 are available now, which makes it possible to determine
the oxygen abundance in a number of H\,{\sc ii} regions in M101 with classic T$_e$ --
method ((O/H)$_{T_{e}}$ abundances). The
(O/H)$_{T_{e}}$ data is sufficient in quantity and quality for an accurate
determination of the value of the oxygen abundance gradient within M101
(Pilyugin 2001b). The radii interior and exterior to the R$^*$ value are
sampled. The (O/H)$_{T_{e}}$ distribution does not show the flattering
at the outer zone of M101 (thick solid line in Fig.\ref{figure:izl5457}a,b).
This is decisive proof that the breaks in the slopes
of (O/H)$_{R_{23}}$ and (O/H)$_{P}$ abundance distributions within the disk of
M101 is an artifact caused by the wrong oxygen abundances in H\,{\sc ii} regions with
galactocentric distances larger than  R$^*$.

The break in the slope of the (O/H)$_{R_{23}}$ abundance gradient in the disk of 
barred spiral galaxy NGC1365 was obtained by Roy \& Walsh (1997). Does the 
break in the slope of the (O/H)$_{R_{23}}$ abundance gradient in NGC1365 
have the same nature as that in NGC5457? Around seventy measurements of the 
strong oxygen lines intensities in H\,{\sc ii} regions in the disk of NGC1365  are 
available (Alloin et al. 1981, Pagel et al. 1979, Roy \& Walsh 1997).
Fig.\ref{figure:izl1365}a shows the radial (O/H)$_{R_{23}}$ abundance
distribution  (the triangles) for H\,{\sc ii} regions in the disk of NGC1365 
derived with Eq.(\ref{equation:ohep}). The galactocentric distances are
normalized to the isophotal radius R$_{25}$ adopted to be equal to 5.61 arcmin 
(de Vaucouleurs et al. 1991). Fig.\ref{figure:izl1365}b shows the radial 
(O/H)$_{P}$ abundance distribution (the filled circles) for H\,{\sc ii} regions 
in the disk of NGC1365 derived with Eq.(\ref{equation:ohp}).
As can be seen in Fig.\ref{figure:izl1365}a,b there is break as in the slope
of the (O/H)$_{R_{23}}$ abundance distribution as in the slope of the (O/H)$_{P}$
abundance distribution at the galactocentric distance R$^*$ $\sim$ 0.7R$_{25}$
although the exact position of the point of break is not beyond question 
due to large scatter in oxygen abundance at any fixed radius.

Comparison of Fig.\ref{figure:izl5457}a and Fig.\ref{figure:izl1365}a shows 
a very important difference between radial abundance distributions in 
the disk of M101 and in the disk of NGC1365. The 
(O/H)$_{R_{23}}$ abundance at the point of bend in the disk of NGC1365 is 
12+log(O/H)$_{R_{23}}$ $\sim$ 8.6 and is appreciable higher than the lower 
boundary (12+log(O/H) $\sim$ 8.2) of the upper branch in
the O/H -- R$_G$ diagram while in the case of M101 the (O/H)$_{R_{23}}$
abundance at the point of bend is around 12+log(O/H)$_{R_{23}}$ $\sim$ 8.3 
and is close to the lower boundary of the upper
branch in the O/H -- R$_G$ diagram.
This high value of (O/H)$_{R_{23}}$ abundance at the point of bend in the disk 
of NGC1365 was a reason why the fact that those H\,{\sc ii} regions belong to the 
upper branch in the O/H -- R$_G$ diagram (and cosequently the validity of the 
R$_{23}$ -- method) was not doubted 
in previous studies. However a difference between values of the (O/H)$_{R_{23}}$
abundances at the point of bend in the disk of M101 and in the disk of NGC1365 
can be easy explained. As it was mentioned above the oxygen
abundance derived with the R$_{23}$ -- method involves a systematic error
depending on the excitation parameter P: the less the value of the
excitation parameter P the more overestimated the value of oxygen abundance
is obtained with the R$_{23}$ -- method (Pilyugin 2001a,b).
In the case of M101 the excitation of the H\,{\sc ii} regions with galactocentric
distances around R$^*$ is moderate P $\sim$ 0.7
(Fig.\ref{figure:izl5457}c) and, as a consequence, the (O/H)$_{R_{23}}$ values
in those H\,{\sc ii} regions are only slightly overestimated. On the contrary,
in the case of NGC1365 the excitation of the H\,{\sc ii} regions with galactocentric
distances around R$^*$ is low P $\sim$ 0.3
(Fig.\ref{figure:izl1365}c) and hence the (O/H)$_{R_{23}}$ values
in those H\,{\sc ii} regions are significantly overestimated. 

At the same time, Fig.\ref{figure:izl1365}b shows that the (O/H)$_{P}$ 
distribution in the disk of NGC1365 flattens at the radius where oxygen 
abundance decreases to around 12+log(O/H) = 8.2. This suggests 
that the breaks in the slopes of (O/H)$_{P}$ and (O/H)$_{R_{23}}$ abundance 
distributions within the 
disk of NGC1365 is an artifact caused by the wrong oxygen abundances in H\,{\sc ii} 
regions with galactocentric distances larger than  R$^*$.
Of course, this statement will be an undisputable fact when the low oxygen 
abundance in the outermost H\,{\sc ii} regions in the disk of NGC1365 will be confirmed 
by the direct determination with the T$_e$ -- method. This statement can be also 
strengthened by consideration of other galaxies.

We have compiled the published measurements
(more than 900 individual measurements) of the strong oxygen lines intensities
in H\,{\sc ii} regions in disks of spiral galaxies and derived the radial (O/H)$_{P}$
abundance distributions. It was found that the oxygen abundances in the
disks of a number of galaxies decreases to 12+logO/H=8.2 within
the isophotal radius. In the disks of ten galaxies the H\,{\sc ii} regions are observed
at the radii exterior to the R$^*$ (where
R$^*$ $\sim$ 0.8R$_{25}$ in NGC300,
R$^*$ $\sim$ 0.5R$_{25}$ in NGC925,
R$^*$ $\sim$ 0.7R$_{25}$ in NGC1365,
R$^*$ $\sim$ 0.9R$_{25}$ in NGC2805,
R$^*$ $\sim$ 0.8R$_{25}$ in NGC3198,
R$^*$ $\sim$ 0.4R$_{25}$ in NGC3319,
R$^*$ $\sim$ 0.8R$_{25}$ in NGC4651,
R$^*$ $\sim$ 0.5R$_{25}$ in NGC5033,
R$^*$ $\sim$ 0.6R$_{25}$ in NGC5457,
R$^*$ $\sim$ 0.6R$_{25}$ in NGC7793).
It is remarkable that many galaxies with reported signs of a bend in the slope 
of the abundance gradient, NGC300, NGC5457, NGC7793 (Vila-Costas \& Edmunds 1992),
NGC3319, NGC5033 (Zaritsky et al. 1994), NGC1365 (Roy \& Walsh 1997) are among 
the galaxies with the H\,{\sc ii} regions observed at the radii exterior to the R$^*$.
If the possibility of a real bend in the slope of the abundance gradient at 
the radius R$^{*}$ in any given galaxy cannot be conclusively excluded, 
it is almost beyond belief that a real bend in the slope of abundance gradient
takes place at the fixed value of oxygen abundance 12+log(O/H) = 8.2 
in all galaxies.
Thus, the fact that galaxies with signs of a bend in abundance gradient
are contained in the list of galaxies with the H\,{\sc ii} regions observed at
the radii exterior to the R$^*$ can be considered as evidence in favour 
that the reported bends are false and reasoned by unjustified use of the
relationship between oxygen abundance and strong line intensities, which is
valid for the H\,{\sc ii} regions of the upper branch of the R$_{23}$ -- O/H diagram
only, to the abundance determination in H\,{\sc ii} regions at radii exterior to
R$^*$, although those H\,{\sc ii} regions do not belong to the upper branch of the
R$_{23}$ -- O/H diagram.

Two galaxies NGC5194 and M81, in which a bend in the slope of the (O/H)$_{R_{23}}$ 
abundance gradient was suspected, will be discussed in the next subsection.
For the disk of galaxy M33, in which the bend in the slope of 
(O/H)$_{R_{23}}$ abundance gradient was reported by Vilchez et al. (1988), 
the compiled data (40 H\,{\sc ii} regions) are reasonably well fitted by a 
single exponential profile.

\subsection{False bends of the "second type"}

The gradient in the excitation parameter P within the disk of M101,
Fig.\ref{figure:izl5457}c, results in the false increase in the
slope of the (O/H)$_{R_{23}}$ abundance gradient, Fig.\ref{figure:izl5457}a.
Hence, if there is a bend in the excitation parameter gradient within the disk
of a galaxy, one can expect that this results in a false bend in the
slope of (O/H)$_{R_{23}}$ abundance gradient. This effect indeed takes place
in the disks of some galaxies. A manifistation of this effect will be 
demonstrated with a galaxy NGC2403.

\begin{figure}
\resizebox{0.80\hsize}{!}{\includegraphics[angle=0]{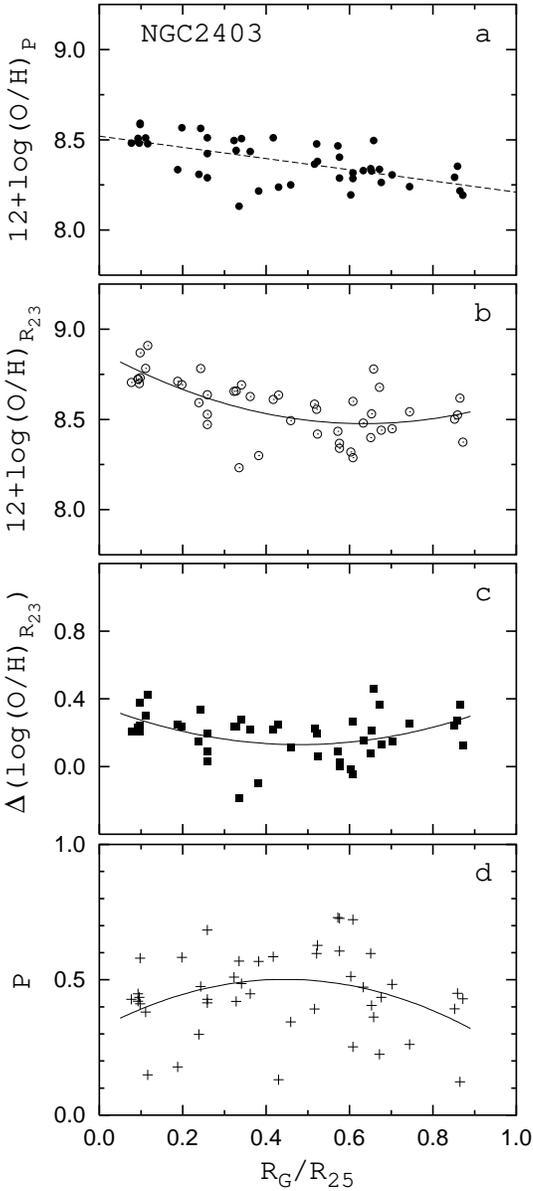}}
\caption{
Gradients in the properties of NGC2403. The galactocentric distances are
normalized to the isophotal radius R$_{25}$.
{\bf a)} The filled circles are (O/H)$_{P}$ abundances in the  H\,{\sc ii} regions,
the dashed line is the best fit (the (O/H)$_{P}$ --  R$_G$ relation).
{\bf b)} The open circles are (O/H)$_{R_{23}}$ abundances in H\,{\sc ii} regions,
the solid curve is the best fit.
{\bf c)} The squares are the deviations of (O/H)$_{R_{23}}$ abundances from
the (O/H)$_{P}$ --  R$_G$ relation. The solid curve is the best fit.
{\bf d)} The pluses are the excitation parameter P in the H\,{\sc ii} regions,
the solid curve is the best fit.
}
\label{figure:izl2403}
\end{figure}

We compiled 47 published measurements of the strong oxygen
line intensities in H\,{\sc ii} regions of the galaxy NGC2403
(Fierro et al. 1986, Garnett et al. 1997, Garnett et al. 1999, McCall et al. 1985,
Smith 1975, van Zee et al. 1998).
Fig.\ref{figure:izl2403}a shows the radial (O/H)$_{P}$ abundance
distribution  (the filled circles) for H\,{\sc ii} regions in NGC2403 derived
with Eq.(\ref{equation:ohp}). The galactocentric distances are
normalized to the isophotal radius R$_{25}$ which was taken to be equal to
11.45 arcmin (de Vaucouleurs et al. 1991).
Fig.\ref{figure:izl2403}b shows the radial (O/H)$_{R_{23}}$ abundance
distribution  (the open circles) for H\,{\sc ii} regions in NGC2403 derived
with Eq.(\ref{equation:ohep}).

Inspection of Fig.\ref{figure:izl2403}a shows that the variation in the logarithm
of (O/H)$_{P}$ abundance with radius can be fitted by a straight line.
At the same time the variation in the logarithm of (O/H)$_{R_{23}}$ abundance
with radius shows evidence of a bend in the slope of the abundance gradient,
Fig.\ref{figure:izl2403}b. The origin of this bend is clearly illustrated in
Fig.\ref{figure:izl2403}c and Fig.\ref{figure:izl2403}d.
Since (O/H)$_{P}$ abundances are as credible as (O/H)$_{T_{e}}$ abundances, 
the (O/H)$_{P}$ --  R$_G$ relation reproduces a real abundance gradient.
Then the deviations of (O/H)$_{R_{23}}$ abundances from the (O/H)$_{P}$ --
R$_G$ relation are errors in (O/H)$_{R_{23}}$ abundances.
Inspection of Fig.\ref{figure:izl2403}c shows that the average value of errors
in (O/H)$_{R_{23}}$ abundances is minimum at galactocentric distance around
0.5R$_{25}$ and inceases with distance from this point.
This behaviour of the average value of errors in (O/H)$_{R_{23}}$ abundances
reflects the behaviour of the average value of the excitation parameter,
Fig.\ref{figure:izl2403}d; the lower the average value of the excitation parameter
the higher the average value of errors in (O/H)$_{R_{23}}$ abundances.

Thus, a false bend in the slope of the (O/H)$_{R_{23}}$ abundance gradient can
appear due to a bend in the excitation parameter gradient within the disk
of a galaxy. The bend in the slope of (O/H)$_{R_{23}}$ abundance 
gradients in the disk of NGC5194 reported by Vila-Costas \& Edmunds (1992) 
is a manifistation of this effect. This effect is also responsible for the 
weak break in the slope of the (O/H)$_{R_{23}}$ abundance gradient in the disk of 
M81 which was indicated by Zaritsky (1992).                                

In general, we did not find a meaningful bend in the radial (O/H)$_{P}$
abundance distributions derived in disks of around 50 spiral galaxies 
investigated. It should be noted however that the number of observed 
H\,{\sc ii} regions in half of the considered galaxies is less than 14. 
Dutil \& Roy (2001) showed that at least 16 H\,{\sc ii} regions are needed 
for a meaningful and robust description of radial abundance gradient in a 
disk galaxy. If one is looking for a break in the gradient of the spatial 
distribution, the sample size required is several times larger. 
Then, the paucity of observational data inhibits the detection of 
a real bend in the radial (O/H)$_{P}$ abundance distribution even if this 
takes place in some galaxies. Thus, we cannot reject the existence of 
a real bend in the radial (O/H)$_{P}$ abundance distribution in galaxies 
at all, but we can conclude that at the present time there is no example of a 
galaxy with an undisputable established bend in the slope of the oxygen abundance 
gradient.

\section{Conclusions}

Spiral galaxies with a reported bend in the slope of gradient in the 
(O/H)$_{R_{23}}$ oxygen abundances (derived with traditionally used R$_{23}$ -- 
method)  were examined. It was found that the false nature of the reported 
bends can be easily explained. Two reasons that result in a false bend in 
the slope of (O/H)$_{R_{23}}$ gradient are indicated.

First, an unjustified use of the relationship between oxygen abundance and 
strong line intensities, constructed for the high-metallicity H\,{\sc ii} regions of 
the upper branch of the R$_{23}$ -- O/H diagram, in the determination of 
oxygen abundance in low-metallicity H\,{\sc ii} regions at the periphery of a galaxy, 
results in wrong (overestimated) oxygen abundances in those H\,{\sc ii} regions, and, 
as a consequence, false bend in the slope of the abundance gradient appears. 
A manifistation of this effect has been demonstrated in a convincing way 
with the galaxy M101. Other galaxies where this effect is expected are listed.

Second, the bend in the excitation parameter gradient within the disk of a 
galaxy results in a false bend in the slope of the (O/H)$_{R_{23}}$ gradient 
since the oxygen abundance derived with the R$_{23}$ -- method involves a 
systematic error depending on the excitation parameter. A manifistation of this 
effect has been demonstrated with the galaxy NGC2403. 

It is concluded that at the present time there is no example of a galaxy with 
an undisputable established bend in the slope of the oxygen abundance gradient.

\begin{acknowledgements}
I thank the anonymous referee for discussion. 
This study was partly supported by the Joint Research Project between 
Eastern Europe and Switzerland (SCOPE) No. 7UKPJ62178, the NATO grant 
PST.CLG.976036, and the Italian national grant delivered by the MURST. 
\end{acknowledgements}

\end{document}